\documentclass[aps,prl,twocolumn,superscriptaddress]{revtex4-1}
\usepackage{graphicx}
\usepackage{subfigure}
\usepackage{epstopdf}
\usepackage{amsmath}
\usepackage{amssymb}
\usepackage{amsfonts}
\usepackage{mathrsfs}
\usepackage{theorem}
\usepackage{bm}
\usepackage{url}
\usepackage[T1]{fontenc}
\usepackage{csquotes}
\MakeOuterQuote{"}
\usepackage{algorithm}
\usepackage{algorithmicx}
\usepackage{algpseudocode}

\usepackage{dcolumn}
\usepackage{color}

\definecolor{ngreen}{rgb}{0.2,0.6,0.2}

\definecolor{ngold}{rgb}{0.7,0.6,0.2}

\newcommand{\be}{\begin{equation}}
	\newcommand{\ee}{\end{equation}}
\newcommand{\ba}{\begin{align}}
	\newcommand{\ea}{\end{align}}
\def\<{\langle}  
\def\>{\rangle}  

\def\eqref#1{\textup{(\ref{#1})}}  
\newcommand{\eref}[1]{Eq.~\textup{(\ref{#1})}}

\newcommand{\fref}[1]{Fig.~\ref{#1}}

\newcommand{\cref}[1]{Conjecture~\ref{#1}}
\newcommand{\Cref}[1]{Conjecture~\ref{#1}}

\begin{document}
	\title{Minimum-consumption discrimination of quantum states via globally optimal adaptive measurements }
	\date{\today}
	\begin{abstract}
Reducing the average resource consumption is the central quest in discriminating non-orthogonal quantum states for a fixed admissible error rate $\varepsilon$. The globally optimal fixed local projective measurement (GOFL) for this task is found to be different from that for previous minimum-error discrimination tasks [PRL 118, 030502 (2017)].
To achieve the ultimate minimum average consumption, here we develop a general globally optimal adaptive strategy (GOA) by subtly using the updated posterior probability, which works under any error rate requirement and any one-way measurement restrictions, and can be solved by a convergent iterative relation. First, under the local measurement restrictions, our GOA is solved to serve as the local bound, which saves 16.6 copies ($24\%$) compared with the previously best GOFL. When the more powerful two-copy collective measurements are allowed, our GOA is experimentally demonstrated to beat the local bound by 3.9 copies ($6.0\%$). By exploiting both adaptivity and collective measurements, our work marks an important step towards minimum-consumption quantum state discrimination.
	\end{abstract}
 \author{Boxuan Tian}
	\affiliation{CAS Key Laboratory of Quantum Information, University of Science and Technology of China, Hefei 230026, P. R. China}
	\affiliation{CAS Center For Excellence in Quantum Information and Quantum Physics, University of Science and Technology of China, Hefei 230026, P. R. China}
 \author{Wen-Zhe Yan}
	\affiliation{CAS Key Laboratory of Quantum Information, University of Science and Technology of China, Hefei 230026, P. R. China}
	\affiliation{CAS Center For Excellence in Quantum Information and Quantum Physics, University of Science and Technology of China, Hefei 230026, P. R. China}
    \author{Zhibo Hou}
	\email{houzhibo@ustc.edu.cn}
	\affiliation{CAS Key Laboratory of Quantum Information, University of Science and Technology of China, Hefei 230026, P. R. China}
	\affiliation{CAS Center For Excellence in Quantum Information and Quantum Physics, University of Science and Technology of China, Hefei 230026, P. R. China}
 \affiliation{Hefei National Laboratory, University of Science and Technology of China, Hefei 230088, People's Republic of China}
	\author{Guo-Yong Xiang}
	\email{gyxiang@ustc.edu.cn}
	\affiliation{CAS Key Laboratory of Quantum Information, University of Science and Technology of China, Hefei 230026, P. R. China}
	\affiliation{CAS Center For Excellence in Quantum Information and Quantum Physics, University of Science and Technology of China, Hefei 230026, P. R. China}
 \affiliation{Hefei National Laboratory, University of Science and Technology of China, Hefei 230088, People's Republic of China}
	\author{Chuan-Feng Li}
	\affiliation{CAS Key Laboratory of Quantum Information, University of Science and Technology of China, Hefei 230026, P. R. China}
	\affiliation{CAS Center For Excellence in Quantum Information and Quantum Physics, University of Science and Technology of China, Hefei 230026, P. R. China}
 \affiliation{Hefei National Laboratory, University of Science and Technology of China, Hefei 230088, People's Republic of China}
	\author{Guang-Can Guo}
	\affiliation{CAS Key Laboratory of Quantum Information, University of Science and Technology of China, Hefei 230026, P. R. China}
	\affiliation{CAS Center For Excellence in Quantum Information and Quantum Physics, University of Science and Technology of China, Hefei 230026, P. R. China}
 \affiliation{Hefei National Laboratory, University of Science and Technology of China, Hefei 230088, People's Republic of China}
	\maketitle
 \begin{figure*}[t]
		\centering\includegraphics[width=0.95\linewidth]{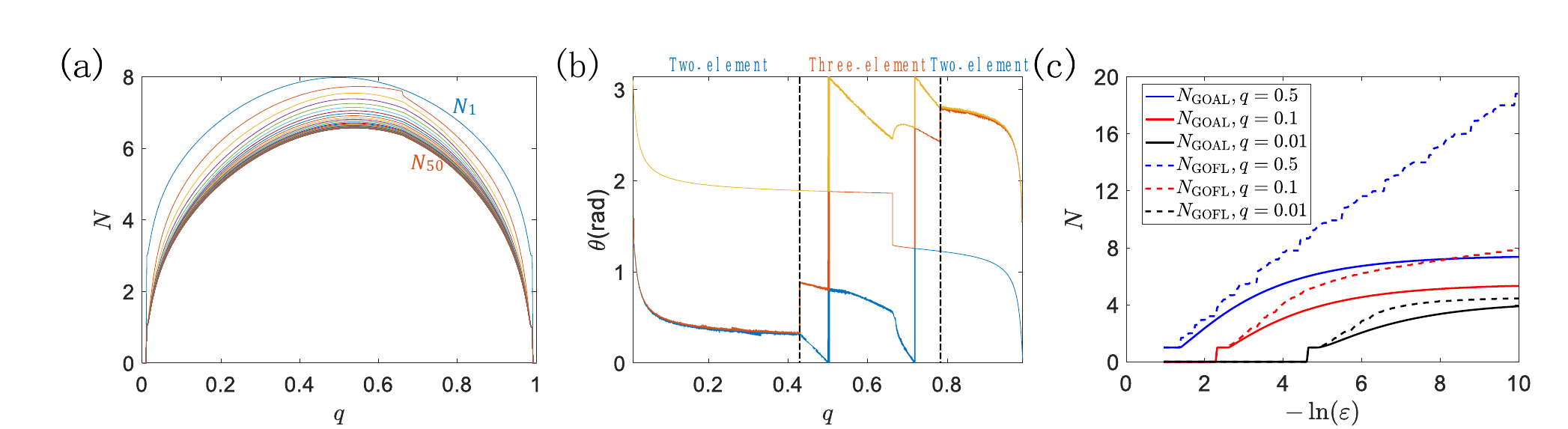}
		\caption{The globally optimal adaptive local strategy. Here the two possible states are $\rho_i =(1-d_i)(\cos x_i|0\rangle+\sin x_i|1\rangle)\left(\cos x_i\langle 0|+\sin x_i\langle 1|\right)+\frac{d_i}{2} I_2$, where $i=0$ or 1, $x_0=\frac{\pi}{12},\ x_1=-\frac{\pi}{12}.$
  In (a) and (b), $\varepsilon=0.01$, $d_0=0.01$ and $d_1=0.001$. In the calculation we discretize $q$ and $\theta$ uniformly into 10000 possible values. The top blue line in (a) is the self-set initial value $N_1(q)$ and the lower lines are the iteration results. After 50 iterations, the results (the bottom orange line) quickly converge to  $N_{\mathrm{GOAL}}(q)$. (b) shows the corresponding numerical result of GOAL strategy $S(q,\theta)$ with nonzero points marked as points. It demonstrates GOAL measurements are either two-element projective measurements or three-element POVM. In (c) GOAL in our work is compared with GOFL in \cite{ref9} at different prior probabilities and error rate requirements and at $d_0=d_1=0$ to show consumption reduction.}
  \label{Fig.1}
	\end{figure*}
 \emph{Introduction.}-- 
Non-orthogonal quantum state discrimination, which can not be perfectly realized with $100\%$ success rate under a limited number of copies, is one of the core problems of quantum information\cite{ref1,ref2,ref4,ref5,ref6,ref7,ref8,ref9,ref10,ref11,ref12,ref13,ref3,ref32,ref30}. Two primary research directions in multiple-copy quantum state discrimination have emerged. One seeks to minimize the average rate of errors given a finite number of copies, a field with a rich history spanning decades \cite{ref1,ref2,ref4,ref5,ref6,ref7,ref8,ref13,ref32,ref30}. The other emerging direction aims to use the minimum average number of copies to achieve a specified error rate requirement $\varepsilon$, thereby conserving quantum resources and enhancing applications like quantum communication \cite{ref14,ref15,ref16} and computation \cite{ref17,ref18,ref19}. Notably, this second direction necessitates different optimal measurement strategies compared to the first, as exemplified by the globally optimal fixed local projective measurement (GOFL) \cite{ref9}, which applies identical local projective measurements to all copies.

Adaptive measurements, which adapt based on evolving knowledge updated with measurement outcomes, offer increased efficiency over fixed measurements \cite{ref2,ref5,ref11}. The most efficient adaptive strategy is \emph{globally optimal adaptive measurement} (GOA), which uses global thinking of all resources rather than the local optimum of each resource to optimize subsequent measurements \cite{ref2,ref5} and thus is optimal among all possible schemes. The GOA strategy in minimum-consumption discrimination is barely explored except for in the small limit of error rate requirements ($-\mathrm{ln}\varepsilon$ is much larger than the quantum relative entropy) \cite{ref10,ref11}.

Collective measurements represent a pivotal concept in quantum state discrimination. They involve simultaneously probing multiple copies of a quantum state, allowing for the extraction of richer information than what individual local measurements on the same number of copies can provide. This property makes collective measurements particularly valuable in scenarios where optimizing resource usage is paramount\cite{ref3,ref13,ref21,ref22,ref24,ref23,ref25,ref33,ref34,ref36}. The power of two-copy collective measurements has been demonstrated experimentally in quantum state tomography\cite{ref21,ref34,ref36}, multi-parameter estimation\cite{ref22,ref24,ref33}, backaction reduction\cite{ref20,ref35}, minimum-error discrimination and orienteering on photonic \cite{ref20,ref21,ref24,ref33,ref35,ref36} and superconducting platforms\cite{ref13,ref22}. These collective strategies harness the quantum entanglement and correlations inherent in multi-copy quantum states to enhance measurement precision and discrimination capabilities, opening up new avenues for resource-efficient quantum information processing.

In this study, we introduce a general GOA strategy that applies under any error rate requirement and any one-way measurement restriction (local or collective) for minimum-consumption discrimination. We present an iterative approach with proven convergence. Initially, for the simplest local measurements, we develop a globally optimal adaptive local strategy (GOAL), serving as the local bound while surpassing GOFL's performance. Subsequently, by incorporating two-copy collective measurements, we advance to a globally optimal adaptive collective strategy (GOAC) that outperforms the local bound. We also experimentally implement GOAC, showcasing its effectiveness in surpassing the local bound for both the small limit of error rate and general error rates.

\emph{Globally Optimal Adaptive Strategy}--
The GOA strategy for minimum-error rate discrimination of two quantum states $\rho_0$ and $\rho_1$ involves not only the prior probability but also the measurement round to do a global search from back to front because of the finite number of copies given \cite{ref2,ref5}. However, in the context of minimum-consumption discrimination, where an infinite number of copies is at our disposal, a unique feature emerges: translational symmetry. This symmetry implies that, regardless of the measurement round, the remaining resources remain infinite and independent, simplifying the GOA strategy considerably.

In its most general form, our GOA strategy takes into account collective measurements involving $n$ copies, with local measurements being a special case when $n=1$. If we consider a set of $n$-copy collective measurements denoted as $\left\{M_k\right\}$, collectively forming a set $D_n$, then the average copy consumption under the GOA strategy can be described as follows:
\begin{equation}
N_{\mathrm{GOA}}(q)=\left\{\begin{array}{c}
0 \quad \text { if } \min \left(q, 1-q\right) \leq \varepsilon,\\
\min \limits_{n,\left\{M_k\right\} \in D_n}[n+\sum\limits_k P_k N_{\mathrm{GOA}}\left(q_k\right)] \text { otherwise. }
\end{array}\right.\label{Eq1}
\end{equation}
Here, $P_k=\mathrm{tr} [M_k(q\rho_0^{\otimes n}+(1-q)\rho_1^{\otimes n})]$ represents the measurement probability of the element $M_k$, and $ q_k=q \mathrm{tr}(M_k\rho_0^{\otimes n})/{P_k}$ corresponds to the posterior probability of $\rho_0$ according to the Bayesian law, updated after obtaining measurement outcome $k$. Importantly, $N_{\mathrm{GOA}}$ depends solely on the prior probability $q$. If $q$ or $1-q$ is already smaller than the error rate $\varepsilon$, no measurements are necessary. Otherwise, we proceed with $n$-copy measurements and update the prior probability using the posterior probability $q_k$. In essence, $N_{\mathrm{GOA}}$ represents the sum of consumed copies in the measurement process and the average copy consumption considering the updated probability $q_k$. It's globally optimal in the sense that it minimizes this sum across all possible local or collective measurements for all values of $q$.

The minimum average consumption $N_{\mathrm{GOA}}$ in \eref{Eq1} is inherently iterative as it appears both sides of the equation, and can be solved iteratively. To initiate this process, we select an initial and feasible average consumption function $N_1\left(q\right)$. Subsequently, we explore all possible measurements to minimize $N_2(q)$ as follows:
$$N_{\mathrm{2}}(q)=\left\{\begin{array}{c}
0 \quad \text { if } \min \left(q, 1-q\right) \leq \varepsilon,\\
\min \limits_{n,\left\{M_k\right\} \in D_n}[n+\sum\limits_k P_k N_{\mathrm{1}}\left(q_k\right)] \text { otherwise. }
\end{array}\right.$$
By following this iterative approach and generating $N_3, N_4, N_5,$ and so on, we can demonstrate its convergence, i.e.(see SM for details),
	\begin{equation}
		\lim _{i \rightarrow+\infty} N_i\left(q\right)=N_{\mathrm{GOA}}\left(q\right)
	\end{equation}
The measurements denoted as $\{M_k\}_{\mathrm{GOA}}(q)$ that achieve $N_{\mathrm{GOA}}$ constitute the globally optimal adaptive strategy. This strategy guides us in selecting adaptive measurements based on updated probability information.

\emph{Globally optimal adaptive local strategy--}
In our current analysis, we narrow our focus within the GOA framework to the most experimentally-friendly local measurements for the discrimination of two possible 2-dimensional real states, $\rho_0$ and $\rho_1$, as a specific example. In this case, all local adaptive strategies can be described by a POVM with rank-one measurement elements $M(\theta)\mathrm{d} \theta$:
	$M(\theta)\mathrm{d} \theta=S_q(\theta)(\cos \theta|0\rangle+\sin \theta|1\rangle)(\cos \theta\langle 0|+\sin \theta\langle 1|) \mathrm{d} \theta,$
	with $\theta \in[0, \pi)$, depending on the prior probability $q$. And $S_q(\theta)$ uniquely determines the POVM and needs to satisfy completeness and semidefinite positivity with four specific conditions
 $\int_0^\pi S_q(\theta) \mathrm{d} \theta=2, \quad \int_0^\pi S_q(\theta) \cos 2 \theta \mathrm{d} \theta=0, \quad \int_0^\pi S_q(\theta) \sin 2 \theta \mathrm{d} \theta=0  \text { \ and \ } S_q(\theta) \geq 0.$ 
  \begin{figure*}[t]
		\centering\includegraphics[width=0.95\linewidth]{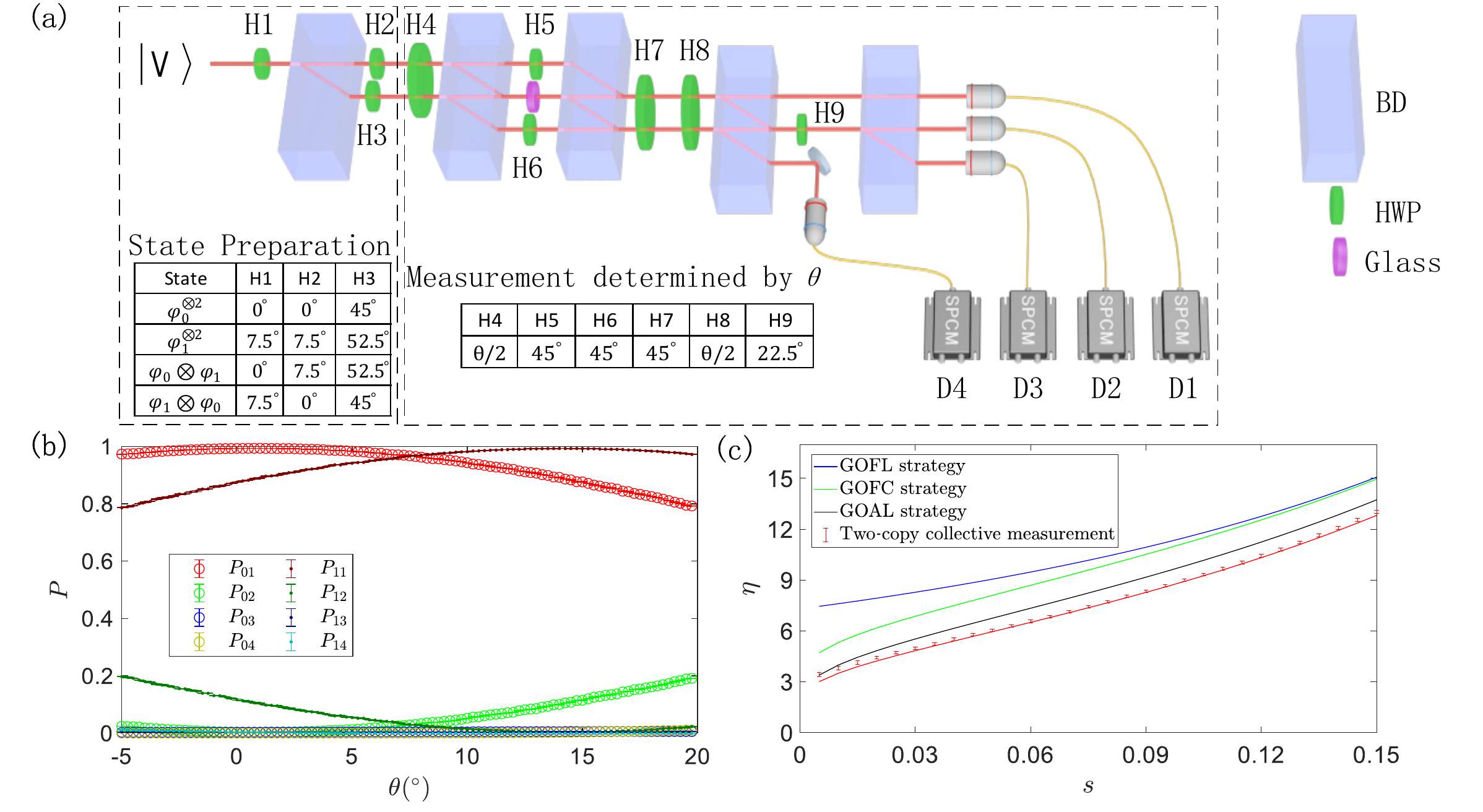}
		\caption{Two-copy collective measurement in the small limit of error rate. (a) shows our experimental setup. In the state preparation module, the sender prepares one of the two mixed states $\rho_0^{\otimes 2}=\left[(1-s)\left(\left|\varphi_0\right\rangle\left\langle\varphi_0\right|\right)+s\left|\varphi_1\right\rangle\left\langle\varphi_1\right|\right]^{\otimes 2}$ and $\rho_1^{\otimes 2}=\left[(1-s)\left(\left|\varphi_1\right\rangle\left\langle\varphi_1\right|\right)+s\left|\varphi_0\right\rangle\left\langle\varphi_0\right|\right]^{\otimes 2}$, where $\left|\varphi_0\right\rangle=|0\rangle$ and $\left|\varphi_1\right\rangle=\cos \frac{\pi}{12}|0\rangle+\sin \frac{\pi}{12}|1\rangle$. In the measurement module, the receiver performs the two-copy collective measurements determined by $\theta$. In (b), a probability distribution is measured for $s=0.05$, where $P_{ik}$ denotes the probability of the measurement outcome $\left|\psi_k\right\rangle\left\langle\psi_k\right|$ on $\rho_i^{\otimes 2}$. Here we take 0.25$^\circ$ as the step length and select 100 values within $-5^{\circ}$ to $20^{\circ}$ for $\theta$. In this article, the half lengths of the error bars are the standard deviation. In (c), the red dots are the minimum value of the ratio $\eta$ we measured for different $s$, where the initial prior probability $q=0.5$. And the red line is the corresponding theoretical value. Compared with GOFL, GOFC and GOAL strategies, the ratio $\eta$ of collective measurements with optimal adaptivity is significantly smaller, which surpasses the local bound.}\label{Fig.2}
	\end{figure*}

All possible forms of the function $S_q(\theta)$ under these constraints compose the set $\mathcal{D}_1$ of local measurements. With this local measurement set $\mathcal{D}_1$, the GOA strategy becomes the globally optimal adaptive local strategy, and \eref{Eq1} is rewritten as
		\begin{equation}
			N_{\mathrm{GOAL}}\left(q\right)=\left\{\begin{array}{c}
				0 \qquad \text { if } \min \left(q, 1-q\right) \leq \varepsilon \\
				1+\min\limits _{S_q(\theta)} \int_0^\pi P_\theta N_{\mathrm{GOAL}}\left(q_\theta\right) d\theta \ \text {otherwise}
			\end{array}\right.\label{Eq3}
		\end{equation}
where $\rho = q \rho_0+\left(1-q\right) \rho_1$, $P_\theta=\operatorname{tr}[M(\theta)\rho]$, $q_\theta=q \operatorname{tr}\left[M(\theta) \rho_0\right]/P_\theta$ is the posterior probability. 
 
 In practice, local POVMs can involve numerous elements, posing a challenge in solving the GOAL strategy outlined in \eref{Eq3}. However, our investigations reveal that GOAL measurements satisfying \eref{Eq3} typically fall into one of two categories: two-element projective measurements or three-element POVMs (see SM for details). Employing a similar iterative approach, we leverage \eref{Eq3} to derive the globally optimal adaptive local strategy denoted as $S_{\mathrm{GOAL}}(q,\theta)$.

In \fref{Fig.1}(a) and 1(b), we provide a visual demonstration of the convergence of the aforementioned iterative procedures to determine $N_{\mathrm{GOAL}}(q)$ and the corresponding two- or three-element measurement strategy. Furthermore, in \fref{Fig.1}(c), we conduct a comparative analysis between the GOAL strategy and the globally optimal fixed local projective measurement as presented in \cite{ref9}. This comparison underscores the efficiency of our approach in the realm of pure state discrimination. Notably, $N_{\mathrm{GOAL}}(q)$ serves as a pivotal local bound, representing the minimum achievable average copy consumption among all conceivable one-way local measurements.

Additionally, for scenarios involving two possible pure states, we found the fitted expression for the globally optimal adaptive local strategy (see SM for details). In situations of $\varepsilon=0$ which has been found to be achievable with finite average number of copies \cite{ref10}, we offer the proof of the expression's correctness, affirming that the average copy consumption achieved by GOAL stands as the definitive lower bound for any prior probability, impervious to challenges from weak measurements or collective measurements \cite{ref10} (see SM for details). However, for scenarios with $\varepsilon>0$, a rigorous mathematical proof remains an open problem.

 \begin{figure*}[t]
		\centering\includegraphics[width=0.95\linewidth]{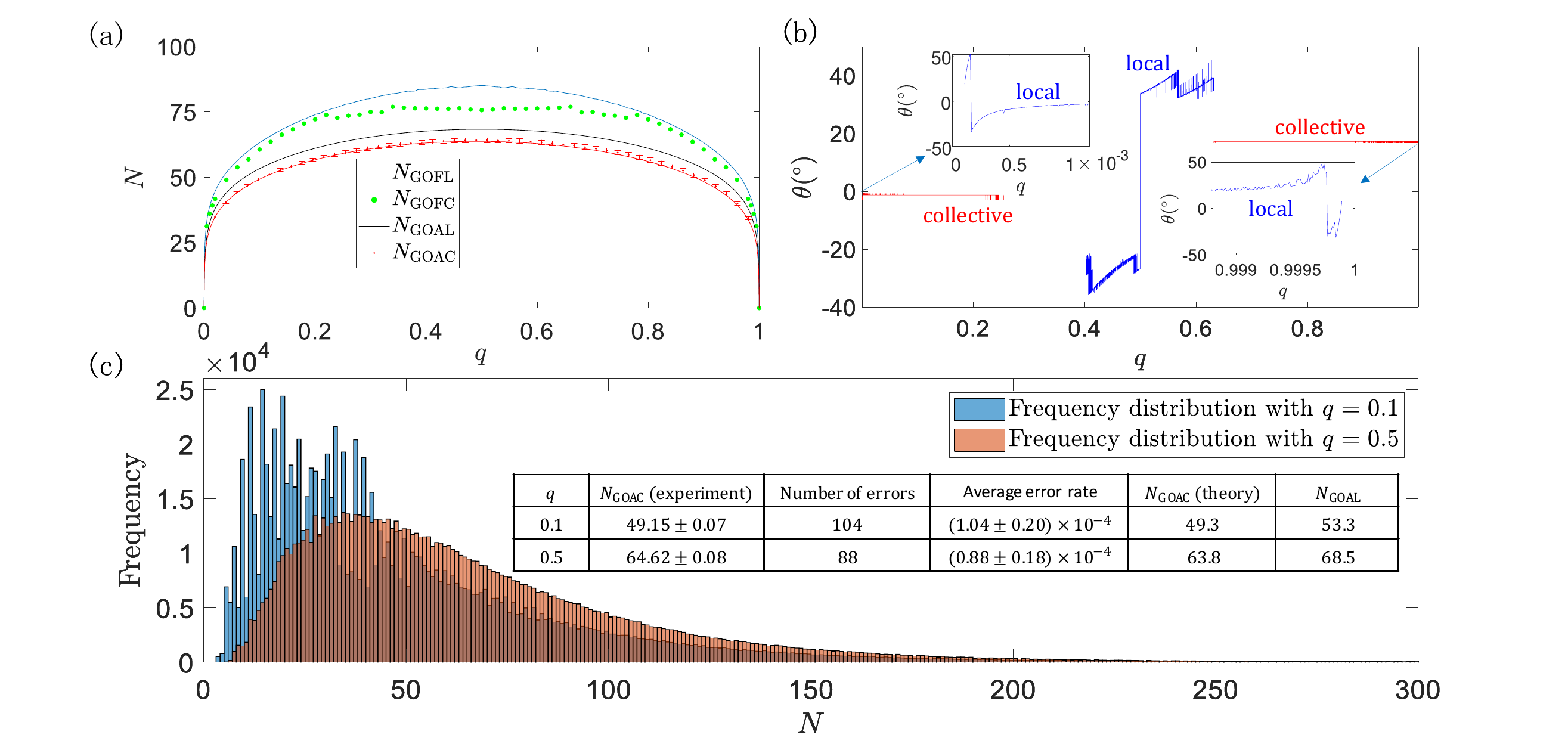}
		\caption{globally optimal adaptive collective strategy for general error rates. Here the two possible states are chosen in the same form as those of \fref{Fig.2}, except for different parameters $s=0.05$ and $\varepsilon=0.0001$. In (a) the red dots represent the average copy consumption of GOAC using the experimental probability distribution of collective measurements while the red lines are from ideal entangling measurements. The measurements in GOAC strategy are plotted in (b) with red lines and blue lines denoting two-copy collective measurements and local measurements, respectively. (c) shows the experimental results of repeating GOAC strategy in (b) for $10^6$ times with two prior probabilities q=0.1 and 0.5, and the statistical chart is a histogram of the frequency distribution of the number of copies consumed in our experiment. The table embedded shows the experimental value of $N_{\mathrm{GOAC}}$ and the actual average error rate. We also listed the theoretical value of $N_{\mathrm{GOAC}}$ and $N_{\mathrm{GOAL}}$ for comparison.}\label{Fig.3}
	\end{figure*}
 
 \emph{Globally optimal adaptive collective strategy--}
 To further reduce the copy consumption and beat the local bound, we incorporate two-copy collective measurements into our GOA strategy to form a globally optimal adaptive collective strategy for mixed state discrimination. For experimental ease, our two-copy collective measurement set is composed of special entangling projective measurements which are easy to implement, whose measurement bases are
$\left|\psi_1\right\rangle=|\theta_{+},\theta_{+}\rangle,
\left|\psi_2\right\rangle=\frac{1}{\sqrt{2}}(|\theta_{+},\theta_{-}\rangle
+|\theta_{-},\theta_{+}\rangle), $
$\left|\psi_3\right\rangle=\frac{1}{\sqrt{2}}(|\theta_{+},\theta_{-}\rangle
-|\theta_{-},\theta_{+}\rangle)$ and
$\left|\psi_4\right\rangle=|\theta_{-},\theta_{-}\rangle$, with $|\theta_{+}\rangle=\cos \theta|0\rangle+\sin \theta|1\rangle$, $|\theta_-\rangle=\sin\theta|0\rangle-\cos \theta|1\rangle$, and $\theta$ ranges from $-5^{\circ}$ to $20^{\circ}$.

In the small limit of $\varepsilon$, the highest efficiency of the device under the prior probability $q$ through adaptive measurement is described by the ratio \cite{ref10,ref11}
 \begin{equation}
\begin{gathered}
\eta \equiv \lim _{\substack{\varepsilon \rightarrow 0 }} \frac{\langle N\rangle_{\min }}{-\ln \varepsilon}
=\frac{q}{\max \limits_{n,\left\{M_k\right\} \in D_n} E_0}+\frac{1-q}{\max \limits_{n,\left\{M_k\right\} \in D_n} E_1}
\end{gathered}
\end{equation}
where $E_0=\frac{1}{n} \sum\limits_k \operatorname{tr}\left(M_k \rho_0^{\otimes n}\right) \ln \frac{\operatorname{tr}\left(M_k \rho_0^{\otimes n}\right)}{\operatorname{tr}\left(M_k \rho_1^{\otimes n}\right)}$ and $E_1=\frac{1}{n} \sum\limits_k \operatorname{tr}\left(M_k \rho_1^{\otimes n}\right) \ln \frac{\operatorname{tr}\left(M_k \rho_1^{\otimes n}\right)}{\operatorname{tr}\left(M_k \rho_0^{\otimes n}\right)}$ are the maximum value of the probabilistic relative entropy divide by the number of copies measured collectively.

Using collective measurements (see \fref{Fig.2}(a)), we experimentally measured the ratio $\eta$ at $q=0.5$ to demonstrate its effectiveness in cases with small $\varepsilon$. In our experiment, we encoded two-copy states using photon polarization and path properties \cite{ref21}. By rotating H4 and H8, we swept $\theta$ for the two-copy collective measurements from $-5^\circ$ to $20^\circ$ and obtained a probability distribution (refer to \fref{Fig.2}(b)). This distribution allowed us to calculate $E_0$ and $E_1$ and, subsequently, determine the $\eta$ ratio for collective measurements with optimal adaptivity. The results, as shown in \fref{Fig.2}(c), not only surpassed GOFL but also outperformed GOAL, the local bound. In \fref{Fig.2}(c), the ratio of globally optimal fixed two-copy collective measurement (GOFC) which is searched from all the fixed strategies through two-copy entangled projective measurements is also presented, underscoring the pivotal role of adaptivity in surpassing the local bound.

For the general error rate $\varepsilon$, 
collective measurements may perform worse than local measurements when $q$ approaches $\varepsilon$ or $1-\varepsilon$ because every local measurement consumes only one copy while every two-copy collective measurement consumes two copies. 
Thus our GOAC strategy also allows local projective measurements with
$\left|\phi_1\right\rangle=\cos \theta|0\rangle+\sin \theta|1\rangle$ and $\left|\phi_2\right\rangle=\sin \theta|0\rangle-\cos \theta|1\rangle$.
where $\theta$ ranges from $0^{\circ}$ to $90^{\circ}$.
The GOAC strategy is then adapted from \eref{Eq1} as
\begin{equation}
\begin{gathered}
N_{\mathrm{G O A C}}\left(q\right)=\min \left\{\min _{\theta \in\left[0^{\circ}, 90^{\circ}\right)}\left[1+\sum_{k=1}^2 P_{k}^L N_{\mathrm{G O A C}}\left(q_{1 k}\right)\right],\right. \\
\left.\min _{\theta \in\left[-5^{\circ}, 20^{\circ}\right]}\left[2+\sum_{k=1}^4 P_{k}^C N_{\mathrm{G O A C}}\left(q_{2 k}\right)\right]\right\}
\end{gathered} \label{Eq5}
\end{equation}where $P_{k}^L=q \operatorname{tr}\left[\left|\phi_k\right\rangle\left\langle\phi_k\right| \rho_0\right]+(1-q) \operatorname{tr}\left[\left|\phi_k\right\rangle\left\langle\phi_k\right| \rho_1\right]$ and $P_{k}^C=q \operatorname{tr}\left[\left|\psi_k\right\rangle\left\langle\psi_k\right| \rho_0^{\otimes 2}\right]+(1-q) \operatorname{tr}\left[\left|\psi_k\right\rangle\left\langle\psi_k\right| \rho_1^{\otimes 2}\right]$ are the measurement probability for local projective measurement and two-copy collective measurements, respectively. And $q_{1k}=\frac{q \operatorname{tr}\left[\left|\phi_k\right\rangle\left\langle\phi_k\right| \rho_0\right]}{P_{\theta1k}}$ and $q_{2k}=\frac{q \operatorname{tr}\left[\left|\psi_k\right\rangle\left\langle\psi_k\right| \rho_0^{\otimes 2}\right]}{P_{\theta2k}}$ are corresponding posterior probability. 

Employing an iterative approach, we determined the GOAC solutions outlined in \eref{Eq5}. At an error rate of $\varepsilon=0.0001$ in \fref{Fig.3}(a), it becomes evident that GOAC outperforms both GOFL, GOFC and GOAL (local bound) strategy across a wide range of prior probabilities. For instance, at q=0.5, GOAC requires an average of 63.8 copies, fewer than both GOAL (68.5 copies), GOFC (75.7 copies) and GOFL (85.1 copies). The measurement characteristics of the GOAC strategy are visually depicted in \fref{Fig.3}(b). Notably, this plot reveals four distinct transition points located at approximately 0.001, 0.4, 0.6, and 0.999. These transitions signify the shift from collective measurements to local measurements and validate that when the prior probability $q$ closely approaches $\varepsilon$, $1-\varepsilon$, or 0.5, local measurements outperform collective measurements.

GOAC experiments are also implemented to demonstrate its resource reduction advantages. Shifting to one-copy state preparation and projective measurement is realized by resetting H2 and H4, respectively. The  GOAC strategy is repeated $10^6$ times and the results with two prior probabilities $q=0.1$ and $0.5$ are shown in \fref{Fig.3}(c). The actual average copy consumption and error rate of the experimental results are close to the theory and the actual average copy consumption of GOAC (49.15 copies and 64.62 copies) clearly beats the local bound in GOAL (53.3 copies and 68.5 copies). 

\emph{Summary.}--
In this work, we introduce the GOA strategy, which has broad applicability across various error rate requirements, pushing forward the theory of minimum-consumption state discrimination. Specifically, under local measurement constraints, we develop a highly efficient GOAL strategy, surpassing all one-way local measurement methods and serving as the local bound.  We then expand our approach to include two-copy collective measurements, resulting in an even more efficient GOAC strategy, which we experimentally implement. Our work connects theory with practical experimentation, underscoring the role of adaptivity and collective measurement in surpassing the local bound. Moreover, our work represents a pioneering experiment that incorporates adaptivity into collective measurements. This achievement opens up exciting new possibilities for advancing the fields of quantum collective measurement \cite{ref3,ref13,ref33,ref34,ref21,ref22,ref24,ref23,ref25,ref20,ref36} and quantum control\cite{ref6,ref26,ref27,ref28,ref29,ref30,ref31,ref2,ref5}.

The work at the University of Science and Technology of China is supported the National Natural Science Foundation of China (Grants Nos. 62222512, 12104439, 12134014, and 11974335), the Innovation Program for Quantum Science and Technology (Grant No. 2021ZD0301203),  the Anhui Provincial Natural Science Foundation (Grant No.2208085J03), USTC Research Funds of the Double First-Class Initiative (Grant Nos. YD2030002007 and YD2030002011) and the Fundamental Research Funds for the Central Universities (Grant No. WK2470000035).

\end{document}


\title{Minimum-consumption discrimination of quantum states via globally optimal adaptive measurements: Supplemental Material}
\maketitle 
In the supplemental material we will discuss following things:

1. Proof: The iterative method can converge to the globally optimal adaptive strategy.

2. The POVM in GOAL strategy has no more than three elements. 

3. The fitted expression of GOAL strategy for pure state discrimination.

4. The globally optimal adaptive local strategy is the best strategy for pure state perfect discrimination even though collective measurements and weak measurements are allowed.

5.The details of the experiment setup.
\section{1. Proof: The iterative method can converge to the globally optimal adaptive strategy.}
Here we prove that the iterative method in the main text converges to the globally optimal adaptive strategy. The iterative method in the main text reads
\begin{equation}
N_{i+1}(q)=\left\{\begin{array}{c}
0 \quad \text { if } \min \left(q, 1-q\right) \leq \varepsilon,\\
\min \limits_{n,\left\{M_k\right\} \in D_n}[n+\sum\limits_k P_k N_i\left(q_k\right)] \text { otherwise. }
\end{array}\right.\label{Eq1}
\end{equation}
The initial $N_1\left(q\right)$ is self-set and realizable. We have $\forall q \in[0,1], N_1\left(q\right) \geq N_{\mathrm{GOA}}\left(q\right)$. As $N_{\mathrm{GOA}}\left(q\right)$ is the minimum average consumption, we have $\forall q \in[0,1]$ and $\forall i \in \mathrm{Z}^{+}, N_i\left(q\right) \geq$ $N_{\mathrm{GOA}}\left(q\right)$. There is at least one strategy to achieve $N_{\mathrm{GOA}}\left(q\right)$, which can be denoted as
$\left\{M_k\right\}_{\mathrm{GOA}}(q)$. 

To facilitate our proof, we use the unknown $\left\{M_k\right\}_{\mathrm{GOA}}(q)$ to define a reference function array $\left\{N_i^{\prime}\left(q\right)\right\}$, which is constructed as	$N_1^{\prime}\left(q\right)=N_1\left(q\right)$  and 
\begin{equation}
 N_{i+1}^{\prime}\left(q\right)=\left\{\begin{array}{c}
0 \quad \text { if } \min \left(q, 1-q\right) \leq \varepsilon,\\
n+\sum\limits_{\left\{M_k\right\}_{\mathrm{GOA}}(q)} P_k N_i^{\prime}\left(q_k\right) \text { otherwise. }
\end{array}\right.\label{Eq2}
\end{equation}
The difference  in \eref{Eq2} is that every round we choose the GOA measurement $\left\{M_k\right\}_{\mathrm{GOA}}(q_k)$ rather than  the optimal measurement to minimize $n+\sum\limits_k P_k N_i\left(q_k\right)$. 
This results in $N_2^{\prime}\left(q\right) \geq N_2\left(q\right)$ because $\left\{M_k\right\}_{\mathrm{GOA}}(q_k)$ is only a possible choice to minimize $N_2\left(q\right)$ and we have $N_1^{\prime}\left(q\right)=N_1\left(q\right)$. To compare between $N_3^{\prime}\left(q\right)$ and $ N_3\left(q\right)$, we define 
\begin{equation}
 N_{i+1}^{a}\left(q\right)=\left\{\begin{array}{c}
0 \quad \text { if } \min \left(q, 1-q\right) \leq \varepsilon,\\
n+\sum\limits_{\left\{M_k\right\}_{\mathrm{GOA}}(q)} P_k N_i\left(q_k\right) \text { otherwise. }
\end{array}\right.\label{Eq3}
\end{equation}
For the same reason as $N_2^{\prime}\left(q\right) \geq N_2\left(q\right)$, we have  $N_3^{a}\left(q\right) \geq N_3\left(q\right)$. Observing the difference in eq 2 and eq 3 and $N_2^{\prime}\left(q\right) \geq N_2\left(q\right)$, we get $N_3^{\prime}\left(q\right) \geq N_3^{a}\left(q\right)$. Then we have $N_3^{\prime}\left(q\right) \geq N_3\left(q\right)$. With the same reasoning, we obtain $\forall i \in \mathrm{Z}^{+}, \forall q \in[0,1], N_i^{\prime}\left(q\right) \geq N_i\left(q\right)$. 

Note that the left-hand-side $N$ terms with index $i+1$ in eqs.1-3 are the functions of prior probability while the right-hand-side terms with index $i$ are functions of posterior probability of the corresponding measurement, implying that terms with larger index are in front of the timeline. Thus $N_i^{\prime}\left(q\right)$ means the average copy consumption realized by doing the globally optimal adaptive strategy in the first $i-1$ steps and if the first $i-1$ steps don't achieve the error rate requirement, doing the strategy corresponding to $N_1\left(q\right)$, we have
\begin{equation}
	\lim _{i \rightarrow \infty} N_i^{\prime}\left(q\right)=N_{\mathrm{GOA}}\left(q\right)\label{Eq4}
\end{equation}
As $\forall i \in \mathrm{Z}^{+}, \forall q \in[0,1], \quad N_{\mathrm{GOA}}\left(q\right) \leq N_i\left(q\right) \leq N_i^{\prime}\left(q\right)$ and $\lim\limits_{i \rightarrow \infty} N_i^{\prime}\left(q\right)=N_{\mathrm{GOA}}\left(q\right)$, finally we have 
\begin{equation}
\lim\limits_{i \rightarrow \infty} N_i\left(q\right)=N_{\mathrm{GOA}}\left(q\right)\label{Eq5}
\end{equation}

\section{2. The POVM in GOAL strategy has no more than three elements}
When searching the globally optimal adaptive local strategy, our task is to calculate the function array $\{N_i\}$ which satisfy following iterative relationship:
\begin{equation}
			N_{\mathrm{i+1}}\left(q\right)=\left\{\begin{array}{c}
				0 \qquad \text { if } \min \left(q, 1-q\right) \leq \varepsilon, \\
				1+\min\limits _{S_q(\theta)} \int_0^\pi P_\theta N_{i}\left(q_\theta\right) d\theta \ \text {otherwise.}
			\end{array}\right.\label{Eq6}
		\end{equation}
and the key problem of the calculation is how to find $S_q(\theta)$ to minimize the integral shown in \eref{Eq6}:
\begin{equation}
	I=\int_0^\pi P_\theta N_{i}\left(q_\theta\right) d\theta\equiv \int_0^\pi S_q(\theta) \operatorname{tr}\left\{\left|\psi_\theta\right\rangle\left\langle\psi_\theta\right|\left[q \rho_0+\left(1-q\right) \rho_1\right]\right\} N_i\left(q_\theta\right) \mathrm{d} \theta\label{Eq7}
\end{equation}
with $\left|\psi_\theta\right\rangle \equiv \cos \theta|0\rangle+\sin \theta|1\rangle$ under the four constraints of $S_q(\theta)$ shown in the main text.

Note that $q_\theta$ is independent of $S_q(\theta)$. Thus for any given $q$, we can separate the unknown $S_q(\theta)$ from the rest in the integral by defining $f(\theta) \equiv S_q(\theta)$ and $g(\theta) \equiv \operatorname{tr}\left\{\left|\psi_\theta\right\rangle\left\langle\psi_\theta\right|\left[q \rho_0+\left(1-q\right) \rho_1\right]\right\} N_i\left(q_\theta\right), \theta \in[0, \pi)$. Then $I=\int_0^\pi f(\theta) g(\theta) d \theta$. Due to the constraint $\int_0^\pi f(\theta) \cos (2 \theta) d \theta=0 \quad$ and $\int_0^\pi f(\theta) \sin (2 \theta) d \theta=0, \forall a, b \in \mathrm{R}$, we have $\int_0^\pi f(\theta) g(\theta) d \theta=\int_0^\pi f(\theta)[g(\theta)+a\cos(2 \theta)+b \sin (2 \theta)] d \theta$. Thus, according to the other two constraints $\int_0^\pi f(\theta)d \theta=2 \quad$ and $f(\theta)\geq0$,
if we can find $a$ and $b$ which satisfy one of two following conditions:

Condition 1: $g(\theta)+a \cos (2 \theta)+b \sin (2 \theta)$ take the minimum value at two points $\theta_1$ and $\theta_2$ and these two points satisfy $\left|\theta_1-\theta_2\right|=\frac{\pi}{2}$.

Condition 2: $g(\theta)+a \cos (2 \theta)+b \sin (2 \theta)$ take the minimum value at three points $\theta_1, \theta_2$ and $\theta_3\left(\theta_1<\theta_2<\right.$ $\theta_3$ ) and these three points satisfy $\theta_1>\theta_2-\frac{\pi}{2}$ and $\theta_2>\theta_3-\frac{\pi}{2}$ and $\theta_3>\theta_1+\frac{\pi}{2}$.

We can immediately obtain the $f(\theta)$ which minimizes the integral :

$$f(\theta)=\left\{\begin{array}{cl}
	\delta\left(\theta-\theta_1\right)+\delta\left(\theta-\theta_2\right) & \text {if} \ a \ \text{and} \ b \text { satisfy \ Condition } 1,\\
	\frac{2 \sin \left[2\left(\theta_2-\theta_1\right)\right] \delta\left(\theta-\theta_3\right)+2 \sin \left[2\left(\theta_3-\theta_2\right)\right] \delta\left(\theta-\theta_1\right)+2 \sin \left[2\left(\theta_1-\theta_3\right)\right] \delta\left(\theta-\theta_2\right)}{\sin \left[2\left(\theta_2-\theta_1\right)\right]+\sin \left[2\left(\theta_3-\theta_2\right)\right]+\sin \left[2\left(\theta_1-\theta_3\right)\right]} & \text {if} \ a \ \text{and} \ b \text { satisfy \ Condition } 2.
\end{array}\right.$$

Here the reason for why $f(\theta)$ can minimize the integral is that it only take positive value in the minimum point of $g(\theta)+a \cos (2 \theta)+b \sin (2 \theta)$, thus for any variations of $f(\theta)$ under the constraints, it is impossible to further reduce the value of integral. Thus it can minimize the value of the integral $\int_0^\pi f(\theta) g(\theta) d \theta=\int_0^\pi f(\theta)[g(\theta)+a\cos(2 \theta)+b \sin (2 \theta)] d \theta$.

For all the actual local measurement devices in the world, no matter how precise, strictly speaking, their measurement angle $\theta$ can only be taken discretely. Thus we can uniformly discretizate the function $g(\theta)$ into an array with very short step distance for $\theta$ and then we can use following method to find $a$ and $b$ which satisfy Condition 1 or Condition 2:
\begin{figure*}[t]
	\centering\includegraphics[width=1\linewidth]{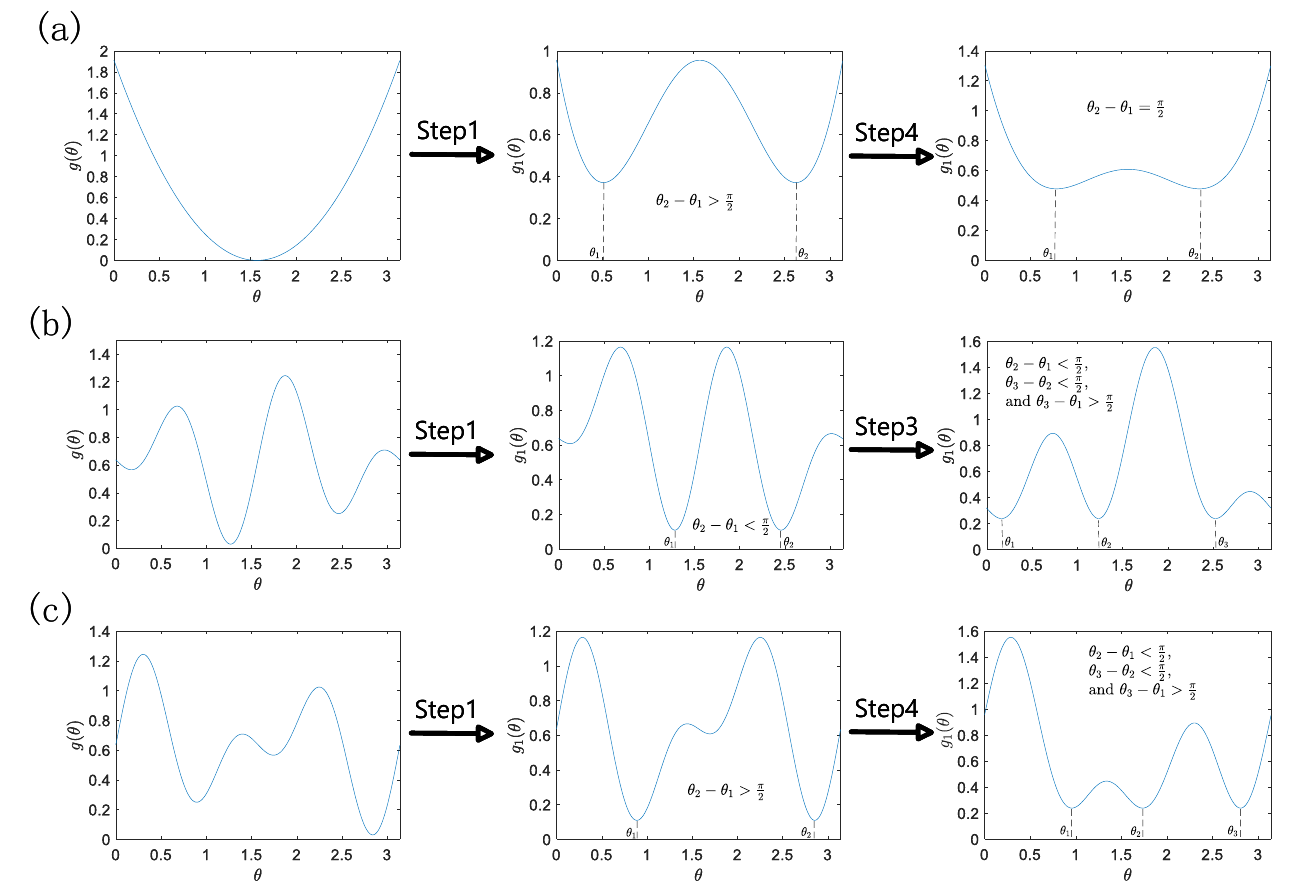}
	\caption{Schematic diagram of the optimal measurement searching algorithm. Here we show three examples. In (a),
 $g(\theta)=\frac{24}{\pi^3}(x-\frac{\pi}{2})^2$, after the first step, the two minimum-value points $\theta_1$ and $\theta_2$ satisfy $\theta_2-\theta_1>\frac{\pi}{2}$, so after the judgement of the second step, we skip the third step and directly do the fourth step. After the fourth step, the two minimum-value points are $\theta_1=\frac{\pi}{4}$ and $\theta_2=\frac{3\pi}{4}$, respectively, which satisfy Condition 1. In (b),$g(\theta)=\frac{1}{\pi}(2+\sin{4x}-\sin{6x})$, after the first step, the two minimum-value points $\theta_1$ and $\theta_2$ satisfy $\theta_2-\theta_1<\frac{\pi}{2}$, so after the judgement of the second step, we should do the third step. After the third step, the three minimum-value points are $\theta_1\approx0.163$, $\theta_2\approx1.236$ and $\theta_3\approx2.528$, respectively, which satisfy Condition 2. So we can finish the searching.
 In (c),
 $g(\theta)=\frac{1}{\pi}(2+\sin{4x}+\sin{6x})$, after the first step, the two minimum-value points $\theta_1$ and $\theta_2$ satisfy $\theta_2-\theta_1>\frac{\pi}{2}$, so after the judgement of the second step, we skip the third step and directly do the fourth step. After the fourth step, the three minimum-value points are $\theta_1\approx0.957$, $\theta_2\approx1.734$ and $\theta_3\approx2.807$, respectively, which satisfy Condition 2.} \label{Fig.1}
\end{figure*}
\begin{enumerate}
\item Step1:Let $g(\theta)+a \cos (2 \theta)+b \sin (2 \theta)$ take the minimum value at two points $\alpha_1$ and $\alpha_2$.\\
Set $a_0=g(\frac{\pi}{2})-g(0)$ and set $g_0(\theta)=g(\theta)+\frac{a_0}{2}\cos(2\theta)$. Then we have $g_0(0)=g_0(\frac{\pi}{2})$.\\
Now  we can set $g_1(\theta)=g_0(\theta)+b_1\sin(2\theta)$. And we can found that when $b\rightarrow-\infty$ the minimum-value point of $g_1$ must in the range $[0, \frac{\pi}{2})$, and when $b\rightarrow+\infty$ the minimum-value point of $g_1$ must in the range $[\frac{\pi}{2},\pi)$. As $g_1(0)=g_1(\frac{\pi}{2})$, it is easy to prove that there must exist $\xi\in R$, when $b_1=\xi$, $g_1(\theta)$ will have at least one minimum-value point in the range $[0,\frac{\pi}{2})$ and will have at least one another minimum-value point in the range $[\frac{\pi}{2},\pi)$. Now we can set $b_1=\xi$ and $g_1(\theta)=g_0(\theta)+b_1\sin(2\theta)$ and then finish Step1. 

\item Step2: In the Step1, we have generated $g_1(\theta)$ which has at least 2 minimum-value points. If we can directly find the minimum-value points which satisfy Condition 1 or Condition 2, we can finish the search.Otherwise,

(1) if $g_1(\theta)$ has only two minimum-value points, set these two points as $\theta_1$ and $\theta_2 .\left(\theta_2>\theta_1\right)$

(2)If $g_1(\theta)$ has more than two minimum-value points, then select $\theta_1$ and $\theta_2$ in the following order:

A. If there exist $\alpha_i$ and $\alpha_j$ which are the minimum-value points of $g_1(\theta)$ and satisfy $\left|\alpha_i-\alpha_j\right|>\frac{\pi}{2}$ we select $\theta_1$ and $\theta_2$ which are the minimum-value points and satisfy $\theta_2-\theta_1=\min \limits_{\alpha_i, \alpha_j,\left|\alpha_i-\alpha_j\right|>\frac{\pi}{2}}\left|\alpha_i-\alpha_j\right|$.

B. If the judgement in A is false, select $\theta_1$ and $\theta_2$ which are the minimum-value points and satisfy
$$
\theta_2-\theta_1=\max _{\alpha_i, \alpha_j}\left|\alpha_i-\alpha_j\right|
$$

\item Step3:
After Step 2, if $\theta_2-\theta_1<\frac{\pi}{2}$, we do as follows:
Conduct transformation $g_1(\theta) \rightarrow g_1(\theta)+A \cdot \cos [2\left(\theta-\frac{\theta_2+\theta_1}{2}\right)]$, where $A=\min\limits_{\theta\in [0,\theta_1)\cup (\theta_2,\pi)}\frac{g(\theta)-g(\theta_1)}{-\cos (\theta-\theta_1-\theta_2)+\cos(\theta_1-\theta_2)}$.
As $g(\theta)$ has been discretized into an array, we can always find a positive $A$.
This transformation will generate at least one more minimum-value point $\theta_3$ in the range $[0,\theta_1)\cup (\theta_2,\pi)$,while $\theta_1$ and $\theta_2$ are still minimum-value points. Then we should do as follows:

A. If there exists minimum-value points which can satisfy Condition 1 or Condition 2, we can finish the search.

B. If all the minimum-value points can't satisfy Condition 1 or Condition 2, do as follows:

B.(1) If there exist two minimum-value points $\theta_i$ and $\theta_j$ which satisfy $\theta_i-\theta_j>\frac{\pi}{2}$, we select $\theta_1^{\prime}$ and $\theta_2^{\prime}$ which are the minimum-value points and satisfy $\theta_2^{\prime}-\theta_1^{\prime}=\min \limits_{\theta_i ,\theta_j,\left|\theta_i-\theta_j\right|>\frac{\pi}{2}}\left|\theta_i-\theta_j\right|$. Then replace $\theta_1$ and $\theta_2$ by $\theta_1^{\prime}$ and $\theta_2^{\prime}$, finish Step 3.

B.(2) If the judgement in B.(1) is false, select $\theta_1^{\prime}$ and $\theta_2^{\prime}$ which are the minimum-value points and satisfy
$\theta_2^{\prime}-\theta_1^{\prime}=\max \limits_{\theta_i, \theta_j}\left|\theta_i-\theta_j\right|$. Then replace $\theta_1$ and $\theta_2$ by $\theta_1^{\prime}$ and $\theta_2^{\prime}$, repeat Step 3.
As every time after repeating Step3, if the judgement in B.(1) is false, the value of $\theta_2-\theta_1$ will increase and be closer to $\frac{\pi}{2}$, Step3 can always finish with a finite number of repetitions.

\item Step4: After doing Step1, 2 and 3, either we have realized the search or $\theta_2-\theta_1>\frac{\pi}{2}$. If the second situation happens, we should do as follows:
Conduct transformation $g_1(\theta) \rightarrow g_1(\theta)-A \cdot \cos [ 2\left(\theta-\frac{\theta_2+\theta_1}{2}\right)]$, where $A=\min\limits_{\theta\in(\theta_1,\theta_2)}\frac{g(\theta)-g(\theta_1)}{\cos (\theta-\theta_1-\theta_2)-\cos(\theta_1-\theta_2)}$.
This transformation will generate at least one more minimum-value point $\theta_3$ in the range $[0,\theta_1)\cup (\theta_2,\pi)$,,while $\theta_1$ and $\theta_2$ are still minimum-value points, then we should do as follows:

A. If these minimum-value points satisfy Condition 1 or Condition 2 , we can finish the search.

B. If the judgement in A is false, for all the minimum-value points $\{\theta_i \}$ we select $\theta_1^{\prime}$ and $\theta_2^{\prime}$ which are the minimum-value points and satisfy  $\theta_2^{\prime}-\theta_1^{\prime}=\min \limits_{\theta_i, \theta_j,\left|\theta_i-\theta_j\right|>\frac{\pi}{2}}\left|\theta_i-\theta_j\right|$.\text Then replace $\theta_1$  and $\theta_2$ by $\theta_1^{\prime} $ and $\theta_2^{\prime}$ , repeat Step 4.

As every time after repeating Step3 and the judgement in A is false, the value of $\theta_2-\theta_1$ will reduce and be closer to  $\frac{\pi}{2}$, Step3 can always finish with a finite number of repetitions. When Step4 finish, we will search the minimum-value points which satisfy Condition 1 or Condition 2.
\end{enumerate}
For clarity, we present three computational examples in \fref{Fig.1}. It is easy to use the above methods to search the globally optimal adaptive local strategy through numerical calculation by the computer. And this method prove that the globally optimal adaptive local strategy which can achieve the local bound only need three-element POVMs and projective measurements.

\section{3. The fitted expression of GOAL strategy for pure state discrimination}
\begin{figure*}[t]
	\centering\includegraphics[width=0.6\linewidth]{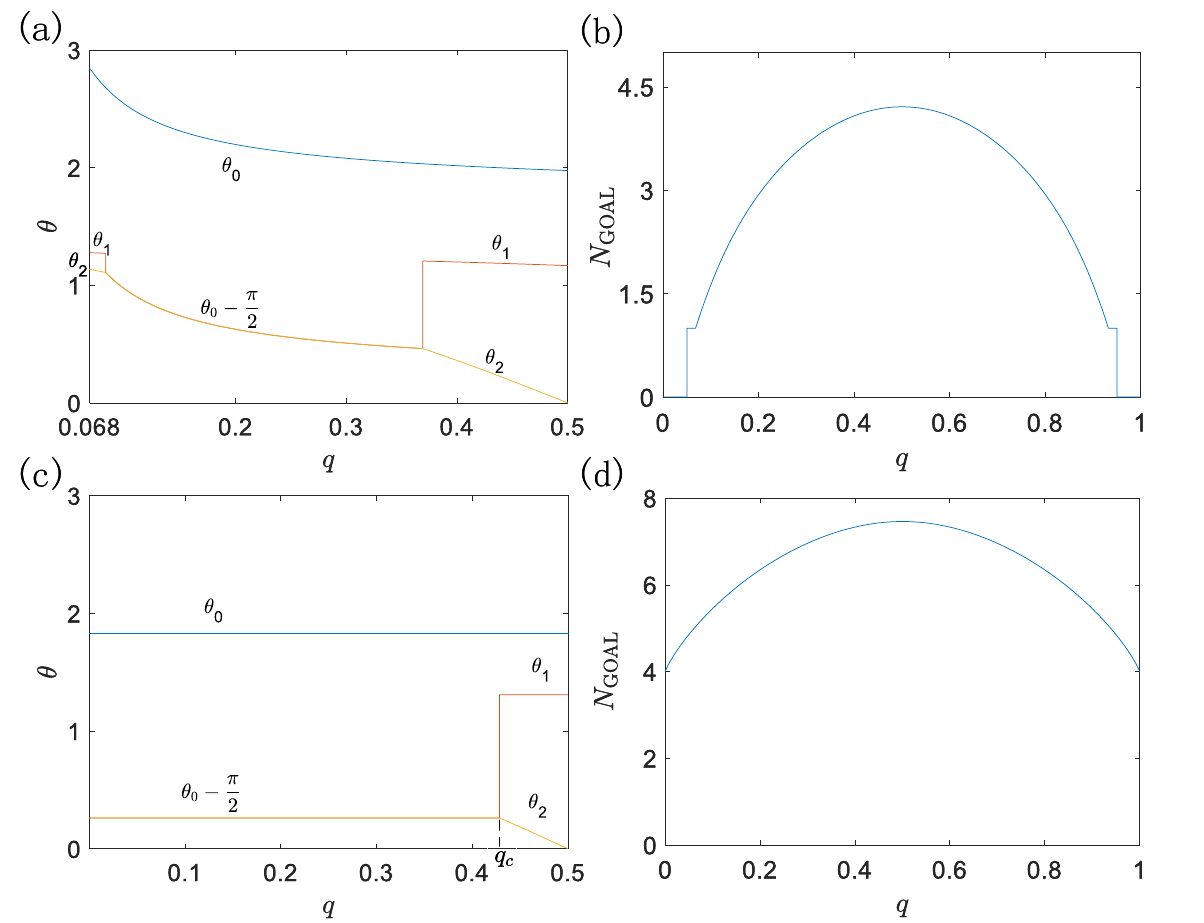}
	\caption{The globally optimal adaptive local strategy for two possible pure states. Here $x=\pi/6$. (a) and (b) show the globally optimal adaptive local measurement when $\varepsilon=0.05$.(c) and (d) show the globally optimal adaptive local measurement when $\varepsilon=0$.In (a) and (c) we only show the area belong to case1, case2 or case3. For the area belong to case4 ($\varepsilon<q\leq\frac{1-\sqrt{1-4 \varepsilon(1-\varepsilon) \cos ^{-2} x}}{2}$), we can realize the requirement with just one more copy \cite{ref1}, and for the area belong to case5 ($q>0.5$), the measurement form is central symmetry about $\left(\frac{1}{2}, \frac{\pi}{2}\right)$. } \label{Fig.2}
\end{figure*}

Through fitting the GOAL for pure state discrimination which are calculated by iterative method, we found the analytic expression of the GOAL for pure state discrimination. Here the two possible pure states are described as
\begin{equation}
\left|\psi_0\right\rangle=\cos \frac{x}{2}|0\rangle+\sin \frac{x}{2}|1\rangle,\left|\psi_1\right\rangle=\cos \frac{x}{2}|0\rangle-\sin \frac{x}{2}|1\rangle\label{Eq8}
\end{equation}
GOAL needs three-element POVM with following three angles $\theta_0(q),\theta_1 (q)$ and $\theta_2(q) $ or projective measurement (See \fref{Fig.2}(a)).

		\begin{scriptsize}
	\begin{equation}
		\begin{aligned}
			& \theta_0=\pi-\frac{1}{2} \arcsin \frac{q-\varepsilon}{\sqrt{\left[\cos (2 x) \varepsilon\left(1-q\right)-(1-\varepsilon) q\right]^2+\left[\varepsilon\left(1-q\right) \sin (2 x)\right]^2}}\\
			& -\frac{1}{2} \arcsin \frac{(1-\varepsilon) q-\cos (2 x) \varepsilon\left(1-q\right)}{\sqrt{\left[\cos (2 x) \varepsilon\left(1-q\right)-(1-\varepsilon) q\right]^2+\left[\varepsilon\left(1-q\right) \sin (2 x)\right]^2}}+\frac{x}{2}, \\
			& \theta_1=-\frac{x}{2}+\frac{1}{2} \arcsin \frac{(1-\varepsilon)\left(1-q\right)-\varepsilon q \cos (2 x)}{\sqrt{\left[(1-\varepsilon)\left(1-q\right)-\varepsilon q \cos (2 x)\right]^2+\left[\varepsilon q \sin (2 x)\right]^2}} \\
			& +\frac{1}{2} \arcsin \frac{1-\varepsilon-q}{\sqrt{\left[(1-\varepsilon)\left(1-q\right)-\varepsilon q \cos (2 x)\right]^2+\left[\varepsilon q \sin (2 x)\right]^2}} ,\\
			& \theta_2=\frac{1}{2} \arcsin \frac{\frac{1}{2}-q}{\sqrt{q^2 \cos ^2 x+\frac{1}{4}-q \cos ^2 x}}
			+\frac{1}{2} \arcsin \frac{\left(\frac{1}{2}-q\right) \cos x}{\sqrt{q^2 \cos ^2 x+\frac{1}{4}-q \cos ^2 x}} .
		\end{aligned}\label{Eq9}
	\end{equation}
        \end{scriptsize}
The process of the optimal strategy is as follows:

case 1:If $q=0.5$, the measurement will be three-element fixed POVM with three angles $\theta_0,\theta_1 \ \text{and} \ \theta_2$. Here $\theta_2=0$. For every copy, if the measurement result is the element with the angle $\theta_2$, the posterior probability is equal to the priori probability. If the measurement result is the element with the angle $\theta_0$ or $\theta_1$, we can make the judgement (the state is $\psi_1$ or $\psi_0$) with the error rate which is equal to $\varepsilon$ and finish the measurement.

case 2:If $0.5>q>\frac{1-\sqrt{1-4 \varepsilon(1-\varepsilon) \cos ^{-2} x}}{2}\ \text{and}\ \theta_0(q)-\frac{\pi}{2}>\theta_2(q)$, we should do adaptive three-element POVM on the first copy with three angles $\theta_0,\theta_1 \ \text{and} \ \theta_2$. If the measurement result is the element with the angle $\theta_2$, the posterior probability will be equal to $0.5$ and the situation turns to case 1 . If the measurement result is the element with the angle $\theta_0$ or $\theta_1$, we can make the judgement (the state is $\psi_1$ or $\psi_0$) with the error rate which is equal to $\varepsilon$ and finish the measurement.

case 3:If $0.5>q>\frac{1-\sqrt{1-4 \varepsilon(1-\varepsilon) \cos ^{-2} x}}{2}\ \text{and}\ \theta_0(q)-\frac{\pi}{2}\leq\theta_2(q)$, we should do adaptive projective measurement with two angles $\theta_0 \ \text{and} \ \theta_0-\frac{\pi}{2}$ on the first copy. 
If the measurement result is the element with the angle $\theta_0$, we can make the judgement (the state is $\psi_1$) with the error rate which is equal to $\varepsilon$ and finish the measurement. If the measurement result is the element with the angle $\theta_0-\frac{\pi}{2}$, the updated posterior probability will increase. If the updated posterior probability still satisfy the condition of case3 we should do adaptive projective measurement on the next copy, otherwise, the situation turns to case 2.

case 4:If $\varepsilon<q\leq\frac{1-\sqrt{1-4 \varepsilon(1-\varepsilon) \cos ^{-2} x}}{2}$, do projective measurement with two eigenvectors of $[q\rho_0-(1-q)\rho_1]$ and then whatever the result is, we can make the judgement and finish the measurement.

case 5:If $q>0.5$, the measurement form is central symmetry about $\left(\frac{1}{2}, \frac{\pi}{2}\right)$ with the part where $q<0.5$.

Then the minimum average consumption can be expressed as follows (See \fref{Fig.2}.1(b)):
\begin{equation}
	N_{\mathrm{GOAL}}(q)=\left\{\begin{array}{c}
		\frac{2}{\lambda \operatorname{tr}\left[|0\rangle\langle 0|\left(\rho_0+\rho_1\right)\right]} \text{  case1,} \\
		1+\lambda \operatorname{tr}\left(\left|\psi_{\theta_2}\right\rangle\left\langle\psi_{\theta_2}\right| \rho\right) N_{\mathrm{GOAL}}(0.5) \text{  case2,} \\
		1+\operatorname{tr}\left(\left|\psi_{\theta_4}\right\rangle\left\langle\psi_{\theta_4}\right| \rho\right) N_{\mathrm{GOAL}}\left(q_{\theta_4}\right) \text{  case3,} \\
		1 \text{  case4,}\\
		N_{\mathrm{GOAL}}(1-q) \text{  case5.}
	\end{array}\right.\label{Eq10}
\end{equation}
where $\theta_4\equiv\theta_0-\frac{\pi}{2}$, $\quad\left|\psi_{\theta_j}\right\rangle \equiv \cos \theta_j|0\rangle+\sin\theta_j|1\rangle$,$\rho=q\left|\psi_0\right\rangle\left\langle \psi_0\left|+\left(1-q\right)\right| \psi_1\right\rangle\left\langle \psi_1\right|$,

$\qquad\lambda=\frac{2 \sin \left[2\left(\theta_1-\theta_0\right)\right]}{\sin \left[2\left(\theta_1-\theta_0\right)\right]+\sin \left[2\left(\theta_2-\theta_1\right)\right]+\sin \left[2\left(\theta_0-\theta_2\right)\right]},\rho_0=\left|\psi_0\right\rangle\left\langle \psi_0\left|, \rho_1=\right| \psi_1\right\rangle\left\langle \psi_1\right|
\text { and } q_{\theta_4}=\frac{q \operatorname{tr}\left(\left|\psi_{\theta_4}\right\rangle\left\langle\psi_{\theta_4}\right| \rho_1\right)}{\operatorname{tr}\left(\left|\psi_{\theta_4}\right\rangle\left\langle\psi_{\theta_4}\right| \rho\right)}$.

Specially, if $\varepsilon=0$, there is one critical point $q_c=\frac{\cos^2x}{1+\cos^2x}$. When $q<q_c$, we should do projective measurement with two angles $\frac{x}{2}+\frac{\pi}{2}$ and $\frac{x}{2}$.When $q_c\leq q\leq1-q$, we should do three-element POVM with three angles $\theta_0=\frac{x}{2}+\frac{\pi}{2}$,
$\theta_1=-\frac{x}{2}+\frac{\pi}{2}$, and $\theta_2=\frac{1}{2} \arcsin \frac{\frac{1}{2}-q}{\sqrt{q^2 \cos ^2 x+\frac{1}{4}-q \cos ^2 x}}+\frac{1}{2} \arcsin \frac{\left(\frac{1}{2}-q\right) \cos x}{\sqrt{q^2 \cos ^2 x+\frac{1}{4}-q \cos ^2 x}} $, and when $1-q_c\leq q$, we should do projective measurement with two angles $-\frac{x}{2}+\frac{\pi}{2}$ and $-\frac{x}{2}$ (See \fref{Fig.2}(c)).For pure states, the minimum average copy consumption is a finite value no larger than $\frac{1}{1-\cos x}$ (this conclusion is the same with [10]), but even if $q \rightarrow 0^{+} \text{or } 1^{-}$, the average copy consumption is no less than $\frac{2}{\sin ^2 x}$ (See \fref{Fig.2}(d)). 


The  reason we infer that it is the globally optimal adaptive local strategy is that 
if we define $g(\theta)=\operatorname{tr}\left|\psi_\theta\right\rangle\left\langle\psi_\theta\right|\left[q\left|\psi_0\right\rangle\left\langle \psi_0|+(1-q)| \psi_1\right\rangle\left\langle \psi_1\right|\right] N_{\mathrm{GOAL}}\left(q_\theta\right), \theta \in[0, \pi)$, where $N_{\mathrm{GOAL}}$ is calculated by \eref{Eq10}, we can find the suitable value of $a$ and $b$ numerically so that for case 3 and case 4, $g(\theta)+a \cos (2 \theta)+b \sin (2 \theta)$ take the minimum value at two points $\theta_0$ and $\theta_0-\frac{\pi}{2}$ and they satisfy Condition 1 and \eref{Eq9}. And for case 1 and case 2, $g(\theta)+a \cos (2 \theta)+b \sin (2 \theta)$ take the minimum value at three points $\theta_0,\theta_1$ and $\theta_2$ and they satisfy Condition 2 and \eref{Eq9}. Thus according to the second part, \eref{Eq10} may be the analytic expression of the average copy consumption realized by GOAL. And how to provide our statement based on numerical calculation evidence with a strict proof is an open problem. However, when $\varepsilon=0$, we have proved that our fitted expression is strictly correct, and the proof is shown in the next section.

\section{4.GOAL is the best strategy for minimum-consumption pure state perfect discrimination even though collective measurements and weak measurements are allowed.}
For pure state minimum-consumption discrimination, it has been known that we can discriminate the two possible states perfectly with finite average number of copies \cite{ref10}. Here we will prove that in the case $\varepsilon=0$, the fitted expression of the GOAL we provide in the last section is the strictly analytic expression and in this case $N_\mathrm{GOAL}(q)$ is the lower bound that can not be surpassed for any error rate requirement and pure state even though collective measurements and weak measurements are allowed. Here the two possible pure states are still expressed by \eref{Eq8}.

First, it is easy to prove that when $\varepsilon=0$, the strategy we provide in the last section is the measurement which maximize the local success rate for perfect discrimination in each measurement round. Thus the minimum failure probability for perfect discrimination realized by measuring $n$ copies collectively can also be calculated according to the measurements shown by \eref{Eq9} and \eref{Eq10}, only $x$ need to be replaced by $y=\arccos(\cos^n(x))$.
That means, the $n$ copies can be described as
\begin{equation}
\left|\psi_0\right\rangle^{\otimes n}=\cos \frac{y}{2}|+\rangle+\sin \frac{y}{2}|-\rangle,\left|\psi_1\right\rangle^{\otimes n}=\cos \frac{y}{2}|+\rangle-\sin \frac{y}{2}|-\rangle\label{Eq17}
\end{equation}
And there will be a critical prior probability $q_c=\frac{\cos^{2n}x}{1+\cos^{2n}x}$. To maximize the success rate for perfect disrimination through measuring these $n$ copies, if $q \in\left(0, q_c\right)$, we should do projective measurement with two angles $\frac{y}{2}+\frac{\pi}{2}$ and $\frac{y}{2}$. If $q \in\left[q_c, 1-q_c\right]$, we should do three-element POVM with three angles $\theta_0=\frac{y}{2}+\frac{\pi}{2}$,
$\theta_1=-\frac{y}{2}+\frac{\pi}{2}$, and $\theta_2=\frac{1}{2} \arcsin \frac{\frac{1}{2}-q}{\sqrt{q^2 \cos ^2 y+\frac{1}{4}-q \cos ^2 y}}+\frac{1}{2} \arcsin \frac{\left(\frac{1}{2}-q\right) \cos y}{\sqrt{q^2 \cos ^2 y+\frac{1}{4}-q \cos ^2 y}} $, and if $q \in\left(1-q_c, 1\right)$, we should do projective measurement with two angles $-\frac{y}{2}+\frac{\pi}{2}$ and $-\frac{y}{2}$.
And we set the minimum failure rate as $P_{n,\mathrm{col}}$ which can be expressed as
\begin{equation}
P_{n, \mathrm{col}}(q,x)=\left\{\begin{array}{c}
2 \sqrt{q(1-q)} \cos ^n x \quad q \in\left[q_c, 1-q_c\right] \\
\min \{q, 1-q\}+\max \{q, 1-q\} \cos ^{2 n} x \quad q \in\left(0, q_c\right) \cup\left(1-q_c, 1\right)
\end{array}\right.\label{Eq11}
\end{equation}

Besides, according to \eref{Eq11}, we can found that $P_{n,\mathrm{col}}$ has the following recursive relation:
\begin{equation}
P_{n, \mathrm{col}}(q,x)=P_{n-1, \mathrm{col}}(q,x)P_{1, \mathrm{col}}(q_{n-1},x)=\prod_{N=1}^{n}P_{1, \mathrm{col}}(q_{N-1},x)\label{Eq16}
\end{equation}
where $q_{N-1}=\min[\frac{\min(q,1-q)}{\max(q,1-q)\cos^{2N-2}x+\min(q,1-q)},0.5]$ represents the smaller one of two possible states' posterior probability. As $P_{1,\mathrm{col}}(q,x)$ is symmetric about $q=0.5$, \eref{Eq16} indicates that the the minimum failure probability $P_{n,\mathrm{col}}$ can be realized just through using the GOAL strategy we shown in the last section.

Thus, during the discrimination process, in every measurement round we can apply one copy and do projective measurement or three-element POVM on it through GOAL strategy. If we do so, the failure rate after $n$ measurement rounds will be equal to $P_{n,\mathrm{col}}$. Thus the average copy consumption will be 
\begin{equation}
\mathrm{N}=1+\sum_{n=1}^{+\infty}P_{n, \mathrm{col}} \equiv \mathrm{N}_{\text {lower bound }}\label{Eq12}
\end{equation}
So through this method we can realize the minimum average copy consumption which reach the lower bound provided in \cite{ref10}.
\section{5. Details of experiment setup}
\subsection{5.1 State Preparation}
In our photonic quantum-walk experiment, the state of each photon is described by the path and polarization\cite{ref21,ref20,ref35,ref36}, i.e.
\begin{equation}
|\psi\rangle=\sum_{p, c} A_{p, c}|p\rangle \otimes|c\rangle\label{Eq13}
\end{equation}
where $p=1,0,-1\cdots$ represents the position and $c\in\{0,1\}$ means the polarized state. Here we set $0\equiv V$ means the vertically polarized state and $1\equiv H$ means the horizontally polarized state.

Now let's consider the state preparation part shown in Fig.2(a) in the main text.
At first, the $V-$polarized photon enter through the path $p=0$. Then it goes through the wave plate H1 which can be seen as a transformation on polarization-encoded states determined by the rotation angle $\alpha_1$ (here $\alpha_1$ is the angle between the optical axis and the vertical direction),
\begin{equation}
U_1=\left(\begin{array}{cc}
\cos (2 \alpha_1) & \sin (2 \alpha_1) \\
\sin (2 \alpha_1) & -\cos (2 \alpha_1)
\end{array}\right)\label{Eq14}
\end{equation}
and the state of the photon will change to 
$|0\rangle\otimes[\cos (2 \alpha_1)|0\rangle+\sin (2 \alpha_1)|1\rangle] $. After encoding the first qubit, the photon passes the beam displacer (BD), the $H$ component is displaced
into path 1 (bottom), the $V$ component is displaced into path -1 (top) and the information of the first qubit will transfer to the path freedom. Finally, the photon will go through H2 and H3 to encode the second qubit.
\begin{figure*}[t]
	\centering\includegraphics[width=0.6\linewidth]{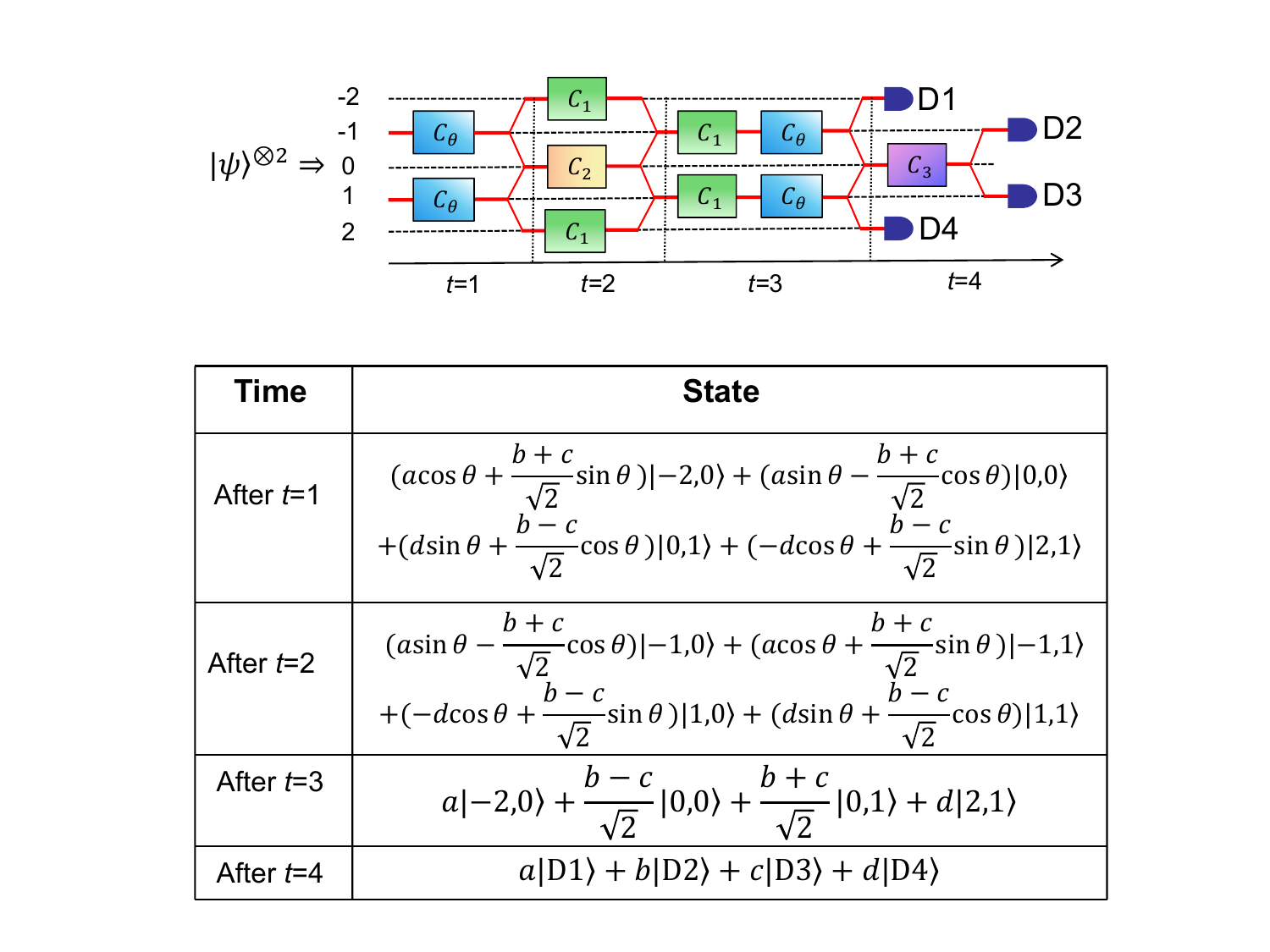}
	\caption{Schematic diagram of quantum walking device.Here the input two-copy state is described as $|\psi\rangle^{\otimes2}=a|\psi_1\rangle+b|\psi_2\rangle+c|\psi_3\rangle+d|\psi_4\rangle$. The optical path diagram shows the scheme of quantum walk and the table shows the photon states after each step. In the path diagram, the four types of transformations are $C_\theta=\left(\begin{array}{cc}\cos \theta & \sin \theta \\ \sin \theta & -\cos \theta\end{array}\right), C_1=\left(\begin{array}{cc}0 & 1 \\ 1 & 0\end{array}\right), C_2=\left(\begin{array}{ll}1 & 0 \\ 0 & 1\end{array}\right) \mathrm{and }\\ C_3=\left(\begin{array}{cc}\frac{\sqrt{2}}{2} & \frac{\sqrt{2}}{2} \\ \frac{\sqrt{2}}{2} & -\frac{\sqrt{2}}{2}\end{array}\right)$, respectively. } \label{Fig.3}
\end{figure*}
As the rotation angles shown in the table of Fig.2(a) in the main text, we can generate different states through rotating the angles of H1, H2 and H3 (set them as $\alpha_1$,$\alpha_2$ and $\alpha_3$, respectively). For example, if we set $\alpha_1=7.5^\circ$,$\alpha_2=7.5^\circ$ and $\alpha_3=52.5^\circ$, the prepared state will be described as
$$|\varphi\rangle_{\text {prepared }}=\left(\cos 15^{\circ}|-1\rangle+\sin 15^{\circ}|1\rangle\right) \otimes\left(\cos 15^{\circ}|0\rangle+\sin 15^{\circ}|1\rangle\right) \equiv\left|\varphi_1\right\rangle^{\otimes 2}.$$
Similarly we can generate $\left|\varphi_0\right\rangle^{\otimes 2}$,$\left|\varphi_0\right\rangle \otimes \left|\varphi_1\right\rangle$ and $\left|\varphi_1\right\rangle \otimes \left|\varphi_0\right\rangle$, and through randomly preparing these pure states, we can generate mixed states. When we want to generate $\rho_0^{\otimes2}$, the probabilities for preparing $\left|\varphi_0\right\rangle^{\otimes 2}$,$\left|\varphi_1\right\rangle^{\otimes 2}$,$\left|\varphi_0\right\rangle \otimes \left|\varphi_1\right\rangle$ and $\left|\varphi_1\right\rangle \otimes \left|\varphi_0\right\rangle$ are $(1-s)^2$,$s(1-s)$,$s(1-s)$ and $s^2$, respectively.When we want to generate $\rho_1^{\otimes2}$, the probabilities for preparing $\left|\varphi_0\right\rangle^{\otimes 2}$, $\left|\varphi_1\right\rangle^{\otimes 2}$, $\left|\varphi_0\right\rangle \otimes \left|\varphi_1\right\rangle$ and $\left|\varphi_1\right\rangle \otimes \left|\varphi_0\right\rangle$ are $s^2$, $s(1-s)$, $s(1-s)$ and $(1-s)^2$, respectively.
\subsection{5.2 Collective Measurement}
Now we will explain why the measurement part of our experiment platform can do two-copy collective measurement. One easy way for analysis is to consider what will happen when the input state is parallel to one of the measurement elements.

For example, if the input state is $\left|\psi_1\right\rangle=\left|\theta_{+}, \theta_{+}\right\rangle\equiv\left(\cos \theta|-1\rangle+\sin \theta|1\rangle\right) \otimes\left(\cos \theta|0\rangle+\sin \theta|1\rangle\right)$ and we set the value of the rotation angles of the wave plates H4, H5, H6, H7, H8 and H9 according to the table shown in Fig.2(a) in the main text, it's easy to find that in both two paths the photon's polarized state will be changed to $H$ by H4. In the next two steps, the two paths of beam will merge and interfere with each other and H7 and H8 will change the merged beam to $H$ polarized state and finally the photon must be received by the single-photon detector D1.
Similarly, it is easy to find that if the input state is $\left|\psi_2\right\rangle=(\left|\theta_{+}, \theta_{-}\right\rangle+\left|\theta_{-}, \theta_{+}\right\rangle)/\sqrt{2}\equiv[\left(\cos \theta|-1\rangle+\sin \theta|1\rangle\right) \otimes\left(\sin \theta|0\rangle-\cos \theta|1\rangle\right)+\left(\sin \theta|-1\rangle-\cos \theta|1\rangle\right) \otimes\left(\cos \theta|0\rangle+\sin \theta|1\rangle\right)]/\sqrt{2}$ or
$\left|\psi_3\right\rangle=(\left|\theta_{+}, \theta_{-}\right\rangle-\left|\theta_{-}, \theta_{+}\right\rangle)/\sqrt{2}\equiv[\left(\cos \theta|-1\rangle+\sin \theta|1\rangle\right) \otimes\left(\sin \theta|0\rangle-\cos \theta|1\rangle\right)-\left(\sin \theta|-1\rangle-\cos \theta|1\rangle\right) \otimes\left(\cos \theta|0\rangle+\sin \theta|1\rangle\right)]/\sqrt{2}$ or $\left|\psi_4\right\rangle=\left|\theta_{-}, \theta_{-}\right\rangle\equiv\left(\sin \theta|-1\rangle-\cos \theta|1\rangle\right) \otimes\left(\sin \theta|0\rangle-\cos \theta|1\rangle\right)$, the photon must be received by D2 or D3 or D4, respectively.

As $\psi_1$,$\psi_2$, $\psi_3$ and $\psi_4$ are orthogonal to each other, for any actually input photon which can be described by the state $|\psi\rangle=\sum_{i=1}^4\left|\psi_i\right\rangle\left\langle\psi_i \mid \psi\right\rangle$, the probability that finally it is received by the detector D$i$ ($i=$ 1 or 2 or 3 or 4) is calculated as
\begin{equation}
P_i=\left|\left\langle\psi_i \mid \psi\right\rangle\right|^2\label{Eq15}
\end{equation}

A more detailed calculation is shown in \fref{Fig.3}, it also proves that our experiment platform can do entangled two-copy collective measurement and the measurement form can be freely transformed by selecting different values of $\theta$.